\documentclass{iau} 
\usepackage{graphicx}
\newcommand{\parcD}[2]{\frac{\partial #1}{\partial #2}}
\newcommand{\dd}{\,\mathrm{d}}
\title[Flare simulations with Flarix] 
{Numerical RHD simulations\\ of flaring chromosphere with {\it {\it Flarix}}}
\author[Heinzel et al.]   
{Petr Heinzel$^1$, Jana Ka\v{s}parov\'{a}$^1$, Michal Varady$^{1,2}$, Marian Karlick\'{y}$^1$ \and 
Zden\v{e}k Moravec$^2$}
\affiliation{$^1$Astronomical Institute of the CAS,
              CZ-25165 Ond\v{r}ejov, Czech Republic \\ email: {\tt
  petr.heinzel@asu.cas.cz} \\
$^2$J.E. Purkyn\v{e} University, Physics Department, \v{C}esk\'{e}
              ml\'{a}de\v{z}e 8, CZ-40096 \'{U}st\'{\i} nad Labem, Czech
              Republic}
\pubyear{2016}
\volume{320}  
\jname{Solar and Stellar Flares and Their Effects on Planets}
\editors{A.G. Kosovichev, S.L. Hawley \& P. Heinzel, eds.}
\begin{document}
\maketitle
\begin{abstract}
{\it Flarix} is a radiation--hydrodynamical (RHD) code
for modeling of the response of the chromosphere to a beam bombardment 
during solar flares.
It solves the set of hydrodynamic
conservation equations coupled with NLTE equations of radiative transfer. 
The simulations are driven by high energy electron beams. We present 
results of the {\it Flarix} simulations of a flaring loop relevant to the
problem of continuum radiation during flares. In particular we focus on
properties of the hydrogen Balmer continuum which was recently detected
by {\it IRIS}.
\end{abstract}

\section{Introduction}
Recent advances in numerical RHD (radiation hydrodynamics) allow to solve complex problems of time
evolution of the solar atmosphere affected by various flare processes (\cite[Allred et al. 2005]{allred2005}, 
\cite[Ka\v{s}parov\'{a} et al. 2009]{ka09}, \cite[Varady et al. 2010]{varady2010}, \cite[Allred et al. 2015]{allred2015}). 
Resulting time-dependent 
atmospheric models (i.e. variations of the temperature, density, ionization and excitation of various species etc.) are then used
as an input for synthesis of spectral lines and continua of atoms and ions under study. One of the 'hot topics' which
attracts a substantial attention is the behavior of the so-called white light flares, and namely the
problem of continuum formation. Recent finding of \cite{hk2014} that
the hydrogen Balmer continuum, previously rarely detected around the Balmer jump, can be easily seen in the flare
spectra taken by {\it Interface Region Imaging Spectrograph IRIS} (\cite[De Pontieu et al. 2014]{bart2014}) in the NUV channel represents a strong motivation for new numerical simulations.
In this paper we first briefly
describe the RHD technique on which our code {\it Flarix} is based and then use the flare simulations to predict the 
time behavior of the Balmer continuum. We then discuss the importance of the Balmer continuum for energy balance
in the flaring chromosphere.

\section{RHD code {\it {\it Flarix}}}
The hybrid radiation--hydrodynamical code {\it Flarix} is based on three
originally auto\-no\-mous, now within {\it Flarix} fully integrated, codes: a 
test-particle (TP) code, a one-dimensional hydrodynamical (HD) 
code and time dependent NLTE radiative transfer code
(for a detailed description of the code see \cite{ka09}
and \cite{varady2010}) . {\it Flarix} is
able to model several processes, which according to contemporary and generally accepted
flare models occur concurrently in flares and play there important roles. 
The transport, scattering and progressive thermalization of the beam electrons
due to Coulomb collisions with particles of ambient plasma in the
magnetized flaring atmosphere and the resulting flare heating
corresponding to the local energy losses of beam electrons is
calculated using an approach proposed by \cite{bai1982} and \cite{karhen1992} based 
on test particles and Monte Carlo method. This approach, fully
equivalent to direct solution of the corresponding Fokker-Planck equation 
(\cite[MacKinnon \& Craig 1991]{mackinnon1991}), provides a flexible 
way to model many various aspects of the beam electron interactions with
the ambient plasma, converging magnetic field in the flare loop or
with additional electric fields (\cite[Varady et~al. 2014]{varady2014}).  
These were proposed in various modifications of the standard CSHKP 
flare model (e.g. \cite[Turkmani et~al. 2006]{turkmani2006}, \cite[Brown
et~al. 2009]{brown2009}, \cite[Gordovskyy \& Browning 2012]
{gordovskyy2012}) or they can be related to the return current propagation 
(\cite[van den Oord 1990)]{oord1990}). 
Owing to TP approach, the detailed distribution function of
beam electrons is known at any instant and position along the flare
loop. This information can be used to calculate a realistic 
distribution of HXR bremsstrahlung sources within the loop, their
position size, spectra and directivity of emanating HXR emission 
(\cite[Moravec et~al. 2013]{moravec2013}, \cite[Moravec et al. 2016]{moravec2016}).

Starting point of {\it Flarix} simulations are parameters of the non-thermal
beam electrons. They are assumed to  obey a single power law in
energy, so their initial spectrum (in units of
\mbox{electrons~cm$^{-2}$~s$^{-1}$~keV$^{-1}$}) is
\begin{eqnarray}\label{eq:powerlaw}
F(E, t) = \left\{\begin{array}{ll}
g(t)F(E) =g(t)\ (\delta-2) 
\frac{F_\mathrm{max}}{E_0^2}\left(\frac{E}{E_0}\right)^{-\delta}
& , \ \ \  \mbox{for} \ \ \  E_0 \le E \le E_1   \\
0  & ,  \  \ \  \mbox{for other} \ E \ .   \\
\end{array} \right.  \ \
\end{eqnarray}
The time dependent electron flux at the loop-top is determined
by $g(t)\in \langle 0,1\rangle$, a function describing the time modulation of the 
beam flux, the maximum energy flux $F_\mathrm{max}$ , i.e. the energy flux of electrons with $E\ge E_0$ at $g(t)=1$,
the low and high-energy cutoffs $E_0$, $E_1$, respectively, and the power-law index $\delta$.
These parameters can be derived either from observations like {\it RHESSI} 
(\cite[Lin et~al. 2002]{lin2002}) or set up arbitrarily.
Another important parameter is the initial pitch angle
distribution of non-thermal electrons $\Theta=\Theta(\vartheta_0)$.  In case the function $\Theta(\vartheta_0)$ is
normalized, the angle dependent initial electron flux is
\begin{equation}
F(E, \vartheta_0, t) =  \Theta(\vartheta_0) F(E, t) \ .
\end{equation}
The initial pitch angle distribution can be chosen by the user.

The 1D one fluid HD part of {\it Flarix} calculates the state and evolution of the
plasma along semicircular magnetic field lines. The following set of equations is solved
\[
\parcD{\rho}{t} + \parcD{}{s}(\rho u) = 0 \qquad
\parcD{\rho u}{t} + \parcD{}{s}(\rho u^2) = - \parcD{P}{s} + F_\mathrm{g}
\]
\begin{equation}
\parcD{E}{t} + \parcD{}{s}(u E) = - \parcD{}{s}(u P) -
\parcD{}{s}{\mathcal{F}}_\mathrm{c}  +
\mathcal{S}\
\end{equation}
where $s$ and $u$ are the position and velocity of plasma along the
fieldlines, respectively, and $\rho$ is the plasma density. The gas pressure $P$, total plasma
energy $E$, and the source term ${\mathcal S}$ are
\begin{equation}
P = n_\mathrm{H} (1 + x +\varepsilon)k_\mathrm{B} T\,, \qquad E=
\frac{P}{\gamma-1} +\frac{1}{2}\rho u^2\ ,
\qquad {\cal S} = {\cal H} - {\cal R} + {\cal Q}\, ,
\end{equation}
where $\gamma= 5/3$ is the ratio of specific heats, $k_\mathrm{B}$ the Boltzmann constant, and $T$
the temperature. The time-dependent hydrogen ionization $x$ is provided by the NLTE code,
$\varepsilon$ accounts for the contribution from metals.

The terms on the right hand side of the system are:
$F_\mathrm{g}$ the component of the gravity force in the parallel direction to
the fieldlines, ${\mathcal F}_\mathrm{c}$ the heat flux calculated according
to the Spitzer classical formula, and
${\mathcal S}$ includes all kinds of heating, i.e. mainly the total
flare heating ${\cal H}$, 
the quiescent heating ${\cal Q}$ assuring stability of the initial unperturbed
atmosphere, and ${\mathcal R}$ the radiative losses. The latter are computed in the NLTE part of the code
by solving the time-dependent radiative-transfer problem in the bottom part of the flaring
loop, for all relevant transitions of hydrogen, CaII and MgII (addition of helium losses is now in progress).
In optically-thin regions of the transition region and corona the loss function of \cite[Rosner et al. (1978)]{rtv78}
is used.

Using the instant values of $T$, $n_\mathrm{H}$, and beam electron energy deposit obtained by the HD
and TP codes, time-dependent NLTE radiative transfer 
for hydrogen is solved in the lower parts of the loop in a 1D plan-parallel approximation.
The hydrogen atom is approximated by a five 
level plus continuum  atomic model. The level populations $n_i$ are determined by the solution of the time-dependent 
system of equations of statistical equilibrium and charge and particle conservation equations
\begin{equation}
\label{ese}
\frac{{\rm d} n_i}{{\rm d} t} = \sum\limits_{j\ne
  i}n_jP_{ji}-n_i\sum\limits_{j\ne i}P_{ij}\ 
\qquad n_{\mathrm e} = n_{\mathrm p} + \varepsilon n_{\mathrm{H}}\, \qquad \sum_{j=1}^5 n_j + n_{\mathrm p} = n_{\mathrm{H}}
,
\end{equation}
where $n_{\mathrm e}$ and $n_{\mathrm p}$ are the electron and proton densities, respectively. $P_{ij}$
contain sums of thermal and non-thermal collisional rates  and radiative rates, the latter 
being preconditioned in the frame of MALI method (\cite[Rybicki \& Hummer 1991]{ryhu91}). 
The NLTE part of {\it Flarix} gives a consistent solution for the non-equilibrium (time-dependent) hydrogen ionization. 

\section{Simulation of a short-duration electron beam heating} 
\begin{figure}[b]
\begin{center}
 \includegraphics[width=8cm]{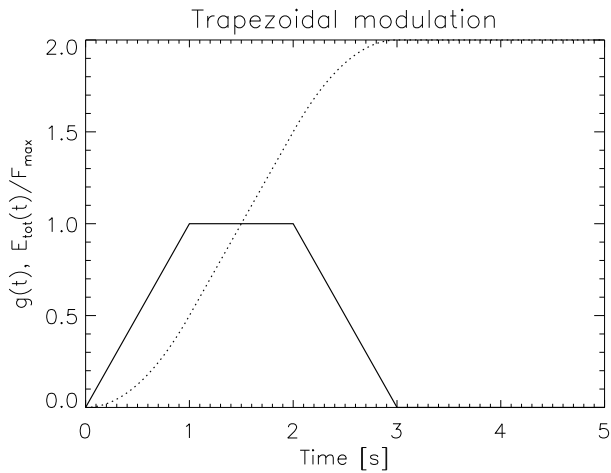} 
 \caption{Time modulation $g(t)$ of the beam flux (solid line), dotted
          line denotes $\int_0^t g(t')\dd t'= E_\mathrm{tot}(t)/ F_\mathrm{max}$.
Total energy deposit is 
$E_\mathrm{tot} = F_\mathrm{max}\int_0^{t_1}g(t) \dd t\ $, where $t_1$ is the duration of the energy deposit.
See also Eqs.~(\ref{eq:powerlaw}).}
\label{fig:timemod}
\end{center}
\end{figure}

Here we present results of a simulation of 
the electron beam heating with a trapezoidal time modulation (see Fig.~\ref{fig:timemod}), 
$\delta=3$, $E_0=20$~keV, $E_1 = 150$~keV, and $F_\mathrm{max}=4.5\times10^{10}$ erg~cm$^{-2}$~s$^{-1}$ which represents a short
beam-pulse heating with a moderate electron beam flux.
As the initial unperturbed atmosphere we used the hydrostatic VAL-C 
atmospheric model of \cite{VAL1981} with a hydrostatic extension
into the transition region and corona.  Response of this atmosphere to
the heating and the temporal evolution of the Balmer line emission 
is discussed in detail in \cite{ka09}, see model H\_TP\_D3 there. In this particular RHD
simulation, a simplified
approach of \cite{pes82} was used to compute chromospheric radiative losses ${\mathcal R}$.

\section{Hydrogen Balmer continuum}

\begin{figure} 
   \centerline{\hspace*{0.015\textwidth}
               \includegraphics[width=0.45\textwidth,clip=]{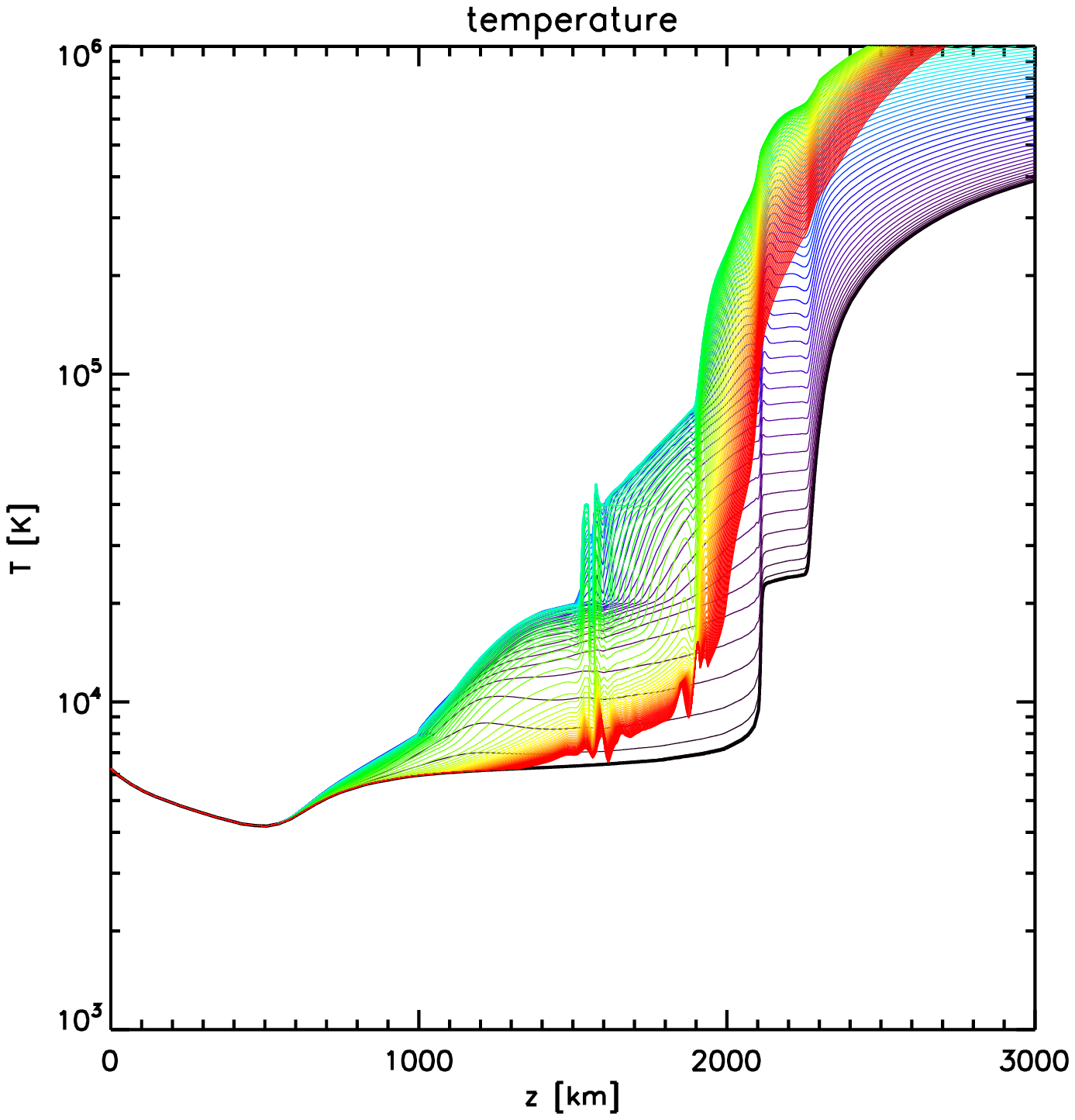}
               \includegraphics[width=0.45\textwidth,clip=]{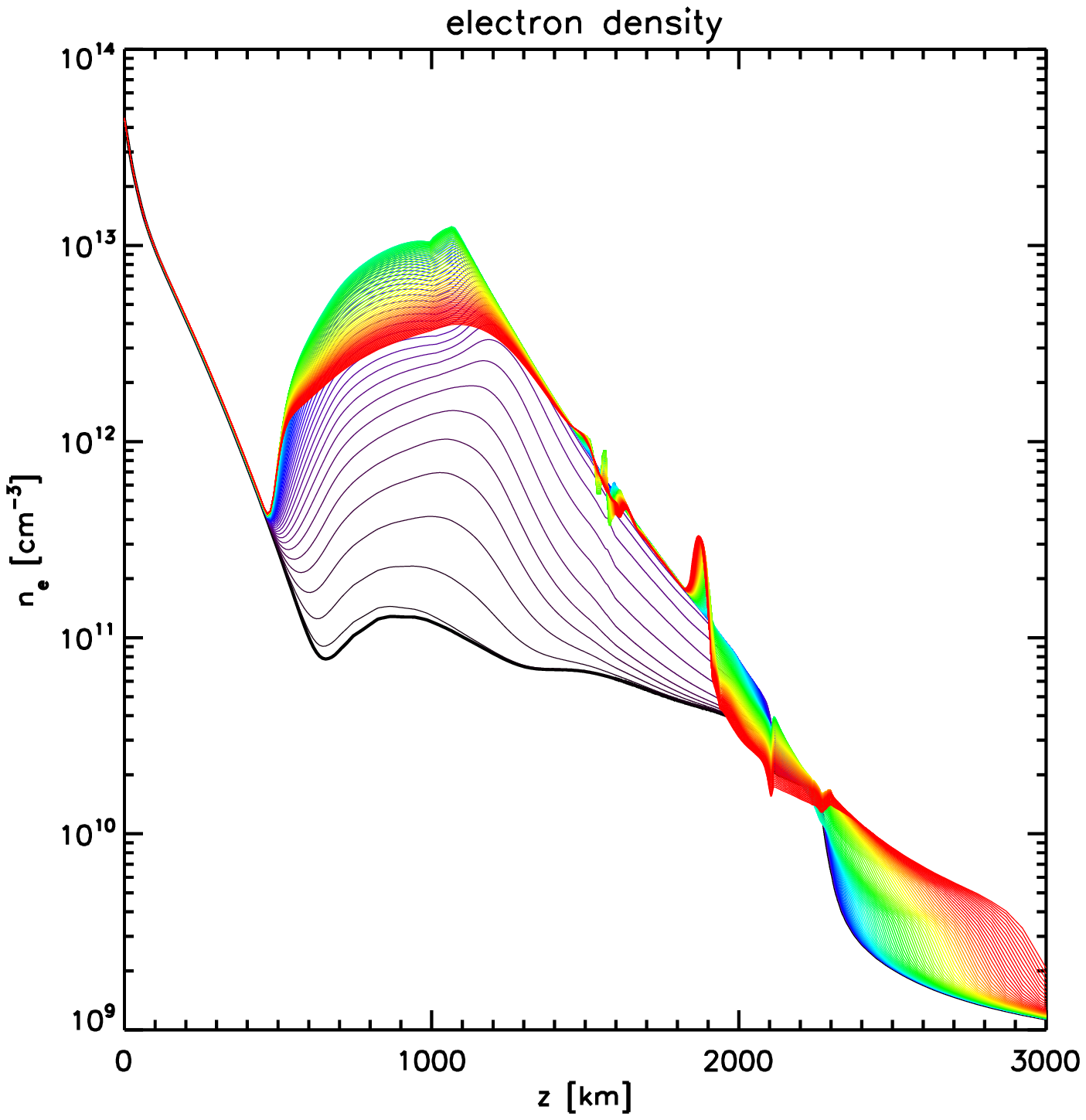}
              }
   \centerline{\hspace*{0.015\textwidth}
               \includegraphics[width=0.45\textwidth,clip=]{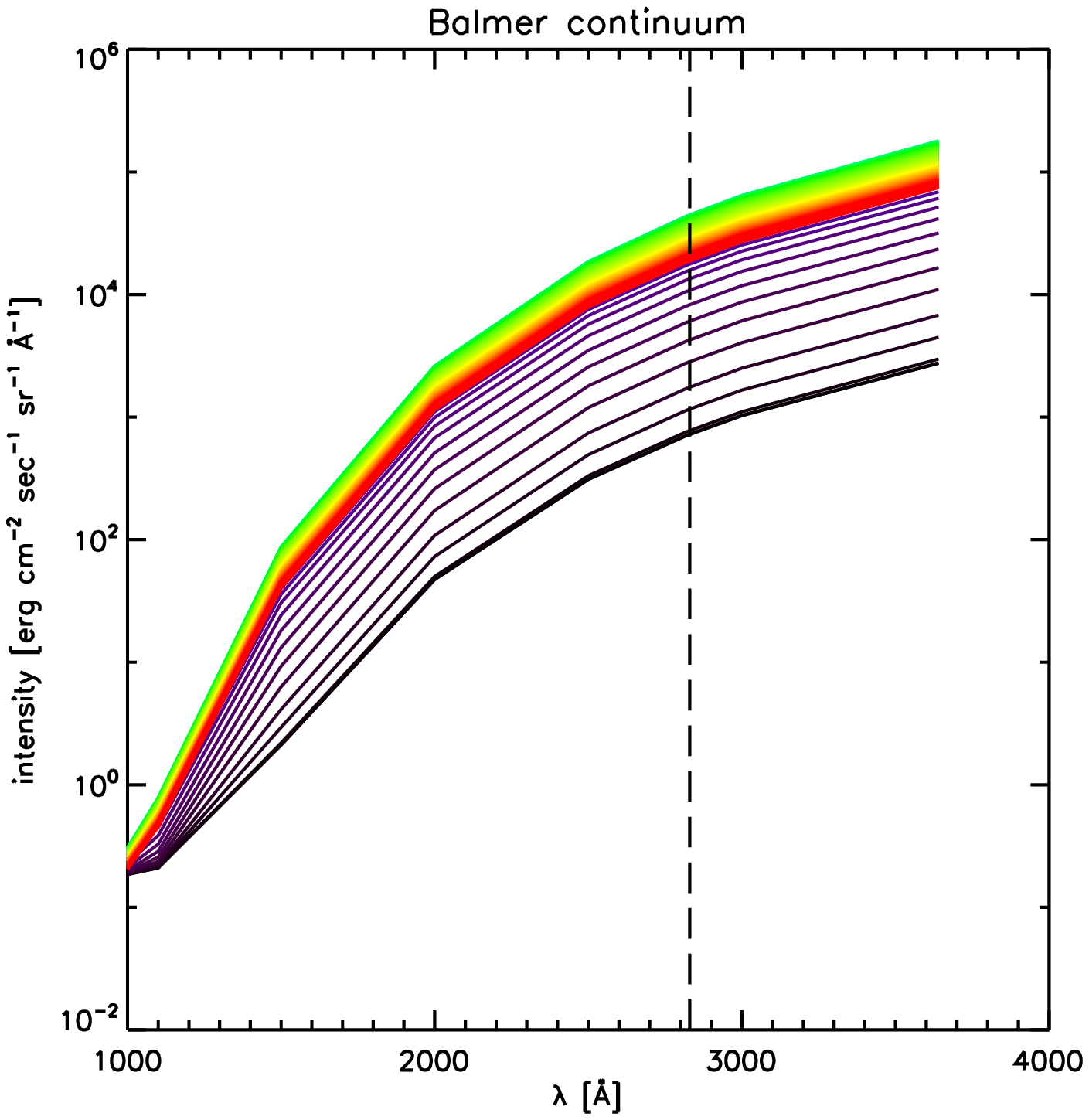}
              }
\caption{Results of {\it Flarix} simulations for the trapezoidal electron beam pulse described in Section 3. The three panels
show, respectively, the time evolution of temperature (top left),
electron density (top right), and the hydrogen
Balmer-continuum spectrum (bottom). In the latter panel we indicate the wavelength used
by {\it IRIS} to detect this continuum (see the vertical dashed line).
The gray-scale coding represents the time evolution from initial stages (black) through the flare
maximum (light gray) up to the end of simulation (dark gray). In the
online version of the paper the time evolution is color coded -- for increasing
time the colors change from black to blue-green-yellow-red.}
   \label{fig:simul}
\end{figure}

\begin{figure}[t]
\begin{center}
 \includegraphics[width=8cm]{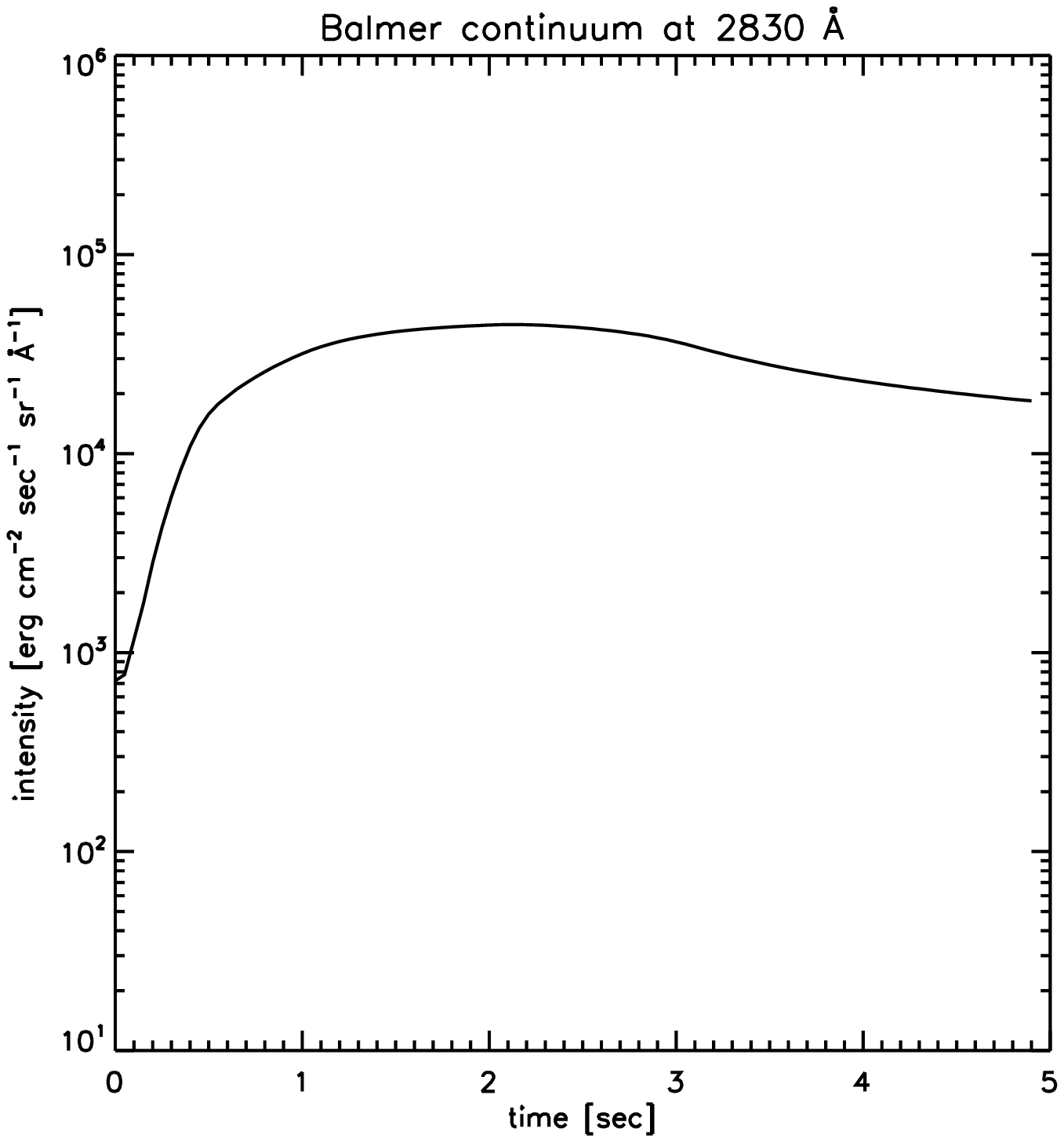} 
 \caption{Simulated time evolution (light-curve) of the hydrogen Balmer continuum intensity at the wavelength selected for
{\it IRIS} detection in the NUV channel (dashed vertical line in Fig. 2).}
\label{fig:lightcurve}
\end{center}
\end{figure}

The time evolution of hydrogen atomic level populations, proton and electron 
densities and plasma temperatures has been used to synthesize the
hydrogen Balmer continuum. 
For this purpose we have performed the formal solution of the
transfer equation and obtained the time evolution of the Balmer continuum 
intensities. The results of {\it Flarix} simulations are shown in
Fig.~\ref{fig:simul}. The time evolution of temperature, primarily due to
the electron beam heating, is shown in the top left panel. The time
evolution of electron density consisting of the non-equilibrium
contribution from hydrogen plus contribution from metals dominating 
around the temperature minimum region is shown in the top right panel.
Around the height of 1000~km the electron density is substantially
enhanced (reaching 10$^{13}$ cm$^{-3}$) which is typical for stronger
flares. 
The bottom panel shows time evolution of the
Balmer-continuum spectrum.
The vertical dashed line is drawn at the wavelength 2830 \AA \, which
corresponds to the NUV spectral window of {\it IRIS} used by
\cite{hk2014} to detect the Balmer continuum. The light curve at
this particular wavelength is then shown in Fig.~\ref{fig:lightcurve}.

\section{Discusion and conclusions}
Our first {\it Flarix}  simulations of time evolution of the hydrogen
Balmer-continuum emission are qualitatively consistent with the {\it IRIS} NUV light
curves obtained by \cite{hk2014} and \cite{khjk2016}. 
They show impulsive intensity rise followed by a gradual decrease. 
Rather slow non-equilibrium hydrogen recombination is perceptible in the 
light curve as expected but a more detailed analysis of this behavior is
needed. However, for the present short trapezoidal electron-beam pulse lasting only a few seconds
(Fig.~\ref{fig:timemod}), the synthetic intensity is lower in comparison with
observations from \cite{hk2014} or \cite{khjk2016}.
This can be due to several reasons. First, the electron-beam flux used in this particular
simulation is almost an order of magnitude lower than that derived from {\it RHESSI} spectra
in \cite{khjk2016}. Second, the Balmer-continuum intensity was found to be increasing with
the boundary pressure at $T$=10$^5$ K (see Table 1 in \cite{khjk2016}) and the pressure
in this simulation is low due to insufficient evaporation.
The beam duration is short and we may 
expect that a long-duration pulse or series of beam pulses will lead to stronger Balmer
continuum, more consistent with the {\it IRIS} observations. Finally, for stronger beams one
should not neglect the return currents which will modify the non-thermal hydrogen excitation
and ionization ({\cite[Karlick\'{y} et al. 2004]{kkh2004}). 
Study of all these aspects is now in progress. 
We have also found
that the chromospheric radiative cooling at the pulse maximum is dominated by
the hydrogen subordinate continua and namely by the Balmer continuum --
for static flare models this was already demonstrated by \cite{avrett86}.
Therefore, this continuum plays a critical role in the energy balance
of the flaring chromosphere, the site where most of the electron-beam 
energy is deposited. Since the Balmer continuum is mostly optically
thin in the flaring chromosphere, the total radiative losses
integrated along the relevant formation heights are directly related to the observed
Balmer-continuum spectral intensity. The {\it IRIS} observations thus provide
an extremely important constraint on the flare energetics. Moreover, 
accepting the idea of backwarming of the flare photosphere, one gets 
directly the amount of radiative energy which should heat the lower
atmospheric layers. This issue is discussed in \cite{khjk2016}
using the {\it IRIS} and optical/infrared
continuum observations. We conclude that the {\it Flarix} simulations 
coupled to broad-band continuum observations should provide
a clue to the long-lasting mystery of the white-light flares.

\vspace{0.4cm}
This project was supported by the EC Program FP7/2007-2013 under the F-CHROMA agreement No. 606862
and by the grant P209/12/0103 of the Czech Funding Agency (GACR).



\end{document}